\documentclass[aps,showpacs,floatfix,superscriptaddress,preprintnumbers]{revtex4-1}
\usepackage{graphicx}
\usepackage{amssymb,amsmath}
\usepackage{textcomp}
\usepackage{epsfig}               
\usepackage{ctable}
\usepackage[english]{babel}
\usepackage{booktabs}

\newcommand{\MeV}{\,{\mathrm{MeV}}}
\newcommand{\GeV}{\,{\mathrm{GeV}}}

\begin{document}

\title{Non extensive thermodynamics for hadronic matter with finite chemical potentials}

\author{Eugenio Meg\'{\i}as}
\affiliation{Grup de F\'{\i}sica Te\`orica and IFAE, Departament de 
F\'{\i}sica, Universitat Aut\`onoma de Barcelona, Bellaterra E-08193 
Barcelona, Spain \\ email: emegias@ifae.es}
\author{D\'ebora P. Menezes}
\affiliation{Departamento de F\'isica, CFM, Universidade Federal de Santa 
Catarina, CP 476, CEP 88.040-900 Florian\'opolis - SC - Brazil \\ email: debora.p.m@ufsc.br}
\author{Airton Deppman}
\affiliation{Instituto de F\'isica, Universidade de S\~ao Paulo - Rua do 
Mat\~ao Travessa R Nr.187 CEP 05508-090 Cidade Universit\'aria, S\~ao Paulo - 
Brazil \\ email: deppman@if.usp.br}
 
\begin{abstract}
The non extensive thermodynamics of an ideal gas composed by bosons
and/or fermions is derived from its partition function for systems
with finite chemical potentials. It is shown that the thermodynamical
quantities derived in the present work are in agreement with those
obtained in previous works when $\mu \le m$. However some
inconsistencies of previous references are corrected when $\mu > m$. A
discontinuity in the first derivatives of the partition function and
its effects are discussed in detail.  We show that at similar
conditions, the non extensive statistics provide a harder equation of
state than that provided by the Boltzmann-Gibbs statistics. 
\end{abstract}

\maketitle

\section{Introduction}

One of the main concerns in the study of ultra-relativistic collisions is the investigation of the quark-gluon plasma (QGP) properties. In this regard the thermodynamical aspects of the plasma is specially interesting due not only to the possibility of studying the deconfinement process but also because it can give important information to other fields, like hydrodynamical models of the QGP, cosmological models of the early Universe and models of massive objects in astrophysics. The non extensive statistics has been applied to a large number of problems since the seminal paper by C. Tsallis in 1988~\cite{Tsallis88}. An updated list of applications and studies on the subject can be found in~\cite{Tsallispage}. In High Energy Physics (HEP) nonextensivity  was introduced by Bediaga, Curado and Miranda (BCM) in 2000~\cite{Bediaga}. In that work the authors used the well-known transverse momentum ($p_T$) distribution from Hagedorn's theory~\cite{Hagedorn65} and formally substituted the exponential function 
by the q-exponential function that appears in the Tsallis statistics. With the new distribution obtained by BCM it is possible to describe the whole $p_T$-distribution measured in HEP experiments.

A large number of works reporting the use of the non extensive 
formalism have been published since the BCM work, all of them showing a good agreement with experiments (see for instance ~\cite{1Alice1,5Alice2,8Alice3,12cms3,4Atlas1}). More recently a non extensive generalization of the Hagedorn's theory~\cite{Hagedorn:1984hz,Agasian:2001bj,Broniowski:2004yh,Tawfik:2004sw,Megias:2012kb,Megias:2012hk,Megias:2013xaa} was developed in~\cite{Deppman12} showing that not only a limiting (or critical) temperature, $T_o$, exists but also that there is an entropic index, $q_o$, which characterizes the hadronic systems at least for the confined regime. Also a new formula for the hadronic mass spectrum in terms of $T_o$ and $q_o$  was derived from that theory and in Ref.~\cite{Lucas} it is shown that this formula can describe quite well the spectrum of known hadronic states with masses from the pion mass up to $\sim 2.5 \,\GeV$. The non extensive self-consistent theory~\cite{Deppman12} imposes much more restrictive tests to the applicability of the Tsallis statistics in HEP and a 
number of 
analysis of experimental data~\cite{CW12,Sena1,Sena2,Lucas} have shown that the theoretical predictions are 
in agreement 
with the experimental findings.

The non extensive thermodynamics of hadronic matter was already
explored for null chemical potential ($\mu$) systems in
Ref.~\cite{Deppman:2012qt} and compared with Lattice-QCD data, showing
a reasonable agreement.  In the present work we extend the
thermodynamics to finite chemical potential systems, which is of
importance in the study of nucleus-nucleus collisions and of
astrophysical objects.  An important class of compact objects are
protoneutron stars. The understanding of their evolution in time from
the moment they are born as remnants of supernova explosions until
they completely cool down to stable neutron stars, has been a matter
of intense investigation.  All sorts of phenomenological equations of
state (EOS), relativistic and non-relativistic ones, have been used to
describe protoneutron star matter. These EOS are normally parameter
dependent and are adjusted so as to reproduce nuclear matter bulk
properties, as the binding energy at the correct saturation density
and incompressibility as well as ground state properties of some
nuclei~\cite{sw,Glen00,Haensel}. The present work provides the
necessary formalism for the investigation of how the non extensive
statistics affects stellar matter.

The paper is organized as follows: in section~\ref{sec:PF} we introduce the partition function and show that it is in agreement with the $p_T$-distribution used in previous works~\cite{Bediaga, Deppman12, CW12, Plastino}; in section~\ref{sec:thermodynamics} we derive the thermodynamical functions of interest and compare with previous results in the literature; in section~\ref{sec:Thermo_had_sys} we establish the phase transition line between confined and deconfined regimes in the $T\times \mu$ diagram and discuss the properties of the EOS of hadronic matter; and finally we present our conclusions in section~\ref{sec:conclusions}.

\section{Partition function for non extensive thermodynamics}
\label{sec:PF}

 We next outline the main formulas necessary for the development and 
application of the non-extensive formalism to hadronic matter.
Consider the exponential function defined as
\begin{equation}
\begin{cases}
 & e_q^{(+)}(x)=[1+(q-1)x]^{1/(q-1)} \qquad\;\;\,\,\,\,\,\,\,\,\,\,\,\,\,\,\,\,\,\,\,\,\,\,\,\,\,, \; x\geq 0\,, \\
 & e_q^{(-)}(x)=\frac{1}{ e_q^{(+)}(|x|)}=[1+(1-q)x]^{1/(1-q)}\qquad\,, \; x<0\,.
\end{cases} \label{eq:eq}
\end{equation}
We define the q-logarithm as
\begin{equation}
\begin{cases}
& \log^{(+)}_q(x)=\frac{x^{q-1}-1}{q-1} \,, \\
&  \log^{(-)}_q(x)=\frac{x^{1-q}-1}{1-q} \,,  
\end{cases} \label{eq:logq}
\end{equation}
which would correspond to the inverse function of the q-exponential if $\log_q(x)$ were defined by $\log^{(+)}_q(x)$ for $x\geq 1$ ( $\log^{(-)}_q(x)$ for $x<1$). In the following we make use of these two definitions of the q-exponential, but do not consider in general their definition regimes in this way. It follows straightforwardly that
\begin{equation}
\begin{cases}
 & \frac{d}{dx}e_q^{(+)}(x)=[e_q^{(+)}(x)]^{2-q} \,,\\
&  \frac{d}{dx}e_q^{(-)}(x)=[e_q^{(-)}(x)]^{q} \,,\\
& \frac{d}{dx}\log^{(+)}_q(x)=x^{q-2} \,,\\
&  \frac{d}{dx}\log^{(-)}_q(x)=x^{-q}\,.
\end{cases}\label{usefull}
\end{equation}

The relations above are used many times in the following and 
specially for the derivation of the identities:
\begin{equation}
 \frac{d}{dx}\log^{(-)}_q\bigg(\frac{ e_q^{(+)}(x)-\xi}{ e_q^{(+)}(x)}\bigg)= \xi \bigg[\frac{1}{e_q^{(+)}(x)-\xi}\bigg]^{q}\,,
\end{equation}
for $x \geq 0$, and
\begin{equation}
 \frac{d}{dx}\log^{(+)}_q\bigg(\frac{ e_q^{(-)}(x)-\xi}{ e_q^{(-)}(x)}\bigg) = \xi\bigg[\frac{1}{e_q^{(-)}(x)-\xi}\bigg]^{2-q}\,,
\end{equation}
for $x < 0$. In these expressions and in the following we take $\xi= \pm 1$ 
for bosons and fermions respectively. We then define the grand-canonical 
partition function for a non extensive ideal quantum gas as
\begin{eqnarray}
 \log\Xi_q(V,T,\mu) &=&
 -\xi V\int \frac{d^3p}{{(2\pi)^3}} \sum_{r=\pm}\Theta(r x)\log^{(-r)}_q\bigg(\frac{ e_q^{(r)}(x)-\xi}{ e_q^{(r)}(x)}\bigg) \,, \label{partitionfunc}
\end{eqnarray}
where $x= \beta( E_p - \mu)$, the particle energy is $E_p = \sqrt{p^2+m^2}$, with $m$ being the hadron mass and $\mu$ the chemical potential, and $\Theta$ is the step function. The partition function for bosons is defined only for the case where $\mu \leq 
m$, therefore the term with $r=-$ in the integrand is applied only for fermions, and it only  contributes if $\mu > m$. In the limit $q \rightarrow 1$ the q-exponential  reduces to the exponential function and the q-logarithm  reduces to the logarithm function. In this limit Eq.~(\ref{partitionfunc})  reduces to the well-known Fermi-Dirac and Bose-Einstein partition functions for fermions and bosons, respectively, and in fact it is the partition function for the ideal quantum gas in Tsallis statistics, as discussed below.

There is an intrinsic gain in the knowledge of the non extensive partition function, since it is closer to the methods of Statistical Mechanics, however in order to go deeper into more fundamental aspects one needs knowledge about nonperturbative QCD that are not available at present.

\section{Thermodynamical functions}
\label{sec:thermodynamics}

To show that the partition function defined in Eq.~(\ref{partitionfunc}) corresponds to the one for a quantum gas in non extensive statistics we derive the occupation number, average number of particles, energy density and the entropy, and show that the results are identical to the ones obtained in Refs~\cite{CW12,Plastino} in the sector $x\ge 0$. However, some differences are found and discussed for $x<0$.

\subsection{Derivation of thermodynamic quantities from the partition function}
\label{subsec:derivation}

The average number of particles can be obtained through the relations
\begin{equation}
     \langle N \rangle =\beta^{-1}\frac{\partial}{\partial \mu} \log \Xi_q \bigg|_\beta\,. \label{eq:avN}
\end{equation}
Using
relations~(\ref{usefull}) it is easy to show that
\begin{equation}
 \langle N \rangle = V \left[ C_{N,q}(\mu,\beta,m) +  \int \frac{d^3p}{(2\pi)^3} \sum_{r=\pm} \Theta(r x) \bigg(\frac{1}{e_q^{(r)}(x) -\xi }\bigg)^{\tilde{q}}  \right]\,, \label{occnumb}
\end{equation}
where 
\begin{equation}
\tilde{q}=
\begin{cases}
& q \qquad\quad\;\;\, \,,\,\,\,x\geq 0\,, \\
& 2-q \qquad \,,\,\,\,x<0\,.
\end{cases}
\end{equation}
The $p$-independent term $C_{N,q}(\mu,\beta,m)$ demands some explanation. The integrand in Eq.~(\ref{partitionfunc}) is a discontinuous function in $x=0$, i.e. in $p = \sqrt{\mu^2-m^2}$. As a consequence, one has to be careful when acting with the derivative with respect to $\mu$. Let us denote by $\sum_{r=\pm}F^{(r)}(p,\mu)$ the integrand in Eq.~(\ref{partitionfunc}), then one has
\begin{eqnarray}
\langle N \rangle  &=& \beta^{-1} \frac{\partial}{\partial\mu} \left[\int_{0}^{\sqrt{\mu^2-m^2}} dp \, F^{(-)}(p,\mu) + \int_{\sqrt{\mu^2-m^2}}^\infty dp \, F^{(+)}(p,\mu) \right]  \label{eq:Ninte}\\
&=&  - \frac{\mu} {\beta \sqrt{\mu^2-m^2}} \left[ F^{(+)}(x=0^+) - F^{(-)}(x=0^-)\right] +  \beta^{-1} \int_0^\infty dp \, \sum_{r=\pm 1} \frac{\partial}{\partial\mu}F^{(r)}(p,\mu) \, \,. \nonumber
\end{eqnarray}
The first term in the last equality produces the contribution
\begin{equation}
C_{N,q}(\mu,\beta,m) = \frac{1}{2\pi^2}\frac{\mu\sqrt{\mu^2-m^2}}{\beta} \frac{2^{q-1} + 2^{1-q} -2}{q-1} \Theta(\mu-m) 
\end{equation}
in Eq.~(\ref{occnumb}), which is non vanishing only when $\mu > m$. Note also that $C_{N,q}(\mu,\beta)$ is vanishing in the Boltzmann-Gibbs limit $q \to 1$, as in this case the integrand in Eq.~(\ref{partitionfunc}) is a continuous function in $p$. From Eq.~(\ref{occnumb}) one gets the average occupation number
\begin{equation}
 n_q^{(+,-)}(p) = \bigg(\frac{1}{e_q^{(+,-)}(x) - \xi}\bigg)^{\tilde{q}}\,, \label{eq:nq}
\end{equation}
where the signals $(+)$ and $(-)$ correspond to $x\geq 0$ and $x<0$, respectively. In the case $x \geq 0$, this result is identical to the one obtained by CMP~\cite{Plastino} and by Cleymans and Worku (CW)~\cite{CW12}. In addition, observe that for $\mu=0$ and $p$ sufficiently high one can write the occupation number as
\begin{equation}
  n_q^{(+)}(p) = [e_q^{(+)}(x)] ^{-q} \,,
\end{equation}
which is exactly the equation used by BCM in Ref.~\cite{Bediaga} to describe the HEP $p_T$-distributions. Thus we conclude that the occupation number derived from the partition function defined in Eq.~(\ref{partitionfunc}) can correctly describe the results obtained in ultra relativistic collisions, corresponding to the regime of high temperature and low baryonic chemical potential. Regarding the case $x < 0$, which is relevant for the regime of high baryonic chemical potential, there are some discrepancies with previous references that will be explained in section~\ref{subsec:comparison}.

The entropy can be obtained through the relation
\begin{equation}
 S = - \beta^2\frac{\partial}{\partial \beta}\bigg(\frac{\log \Xi_q}{\beta}\bigg) \bigg|_\mu \,, \label{eq:Sthermo}
\end{equation}
resulting
\begin{eqnarray}
 S &=& V\int \frac{d^3p}{(2\pi)^3} \sum_{r=\pm} \Theta(r x) \bigg[ - [\bar{n}_q^{(r)}(x)]^{\tilde{q}} \log_q^{(-r)}\left(\bar{n}_q^{(r)}(x)\right) + \xi [1 + \xi \bar{n}_q^{(r)}(x) ]^{\tilde{q}} \log_q^{(-r)}\left(1 + \xi \bar{n}_q^{(r)}(x)\right) \bigg]\,,  \label{eq:S}
\end{eqnarray}
where we have defined $\bar{n}_q^{(r)}(x) \equiv [n_q^{(r)}(x)]^{1/\bar{q}}$. This result is identical to the CMP entropy defined in Ref.~\cite{Plastino}, either for $r=+$ or $r=-$, but written in terms of q-log functions and for bosons/fermions, as it can be easily checked. Finally, for the sake of completeness, we show also the result for the average energy. It can be computed from the relation
\begin{equation}
\langle E \rangle = -\frac{\partial}{\partial\beta} \log \Xi_q \bigg|_\mu + \frac{\mu}{\beta} \frac{\partial}{\partial\mu} \log \Xi_q \bigg|_\beta\,,
\end{equation}
and the result is
\begin{equation}
 \langle E \rangle = V \left[ C_{E,q}(\mu,\beta,m) +  \int \frac{d^3p}{(2\pi)^3} \sum_{r=\pm} \Theta(r x) E_p \bigg(\frac{1}{e_q^{(r)}(x) -\xi }\bigg)^{\tilde{q}}  \right]\,, \label{occE} 
\end{equation}
with $C_{E,q}(\mu,\beta,m) = \mu \, C_{N,q}(\mu,\beta,m)$. We have verified the thermodynamic consistency of these expressions by checking that $\left( \frac{\partial S}{\partial E}\right)_{V,N} = \beta$. If the $p$-independent terms $C_{N,q}(\mu,\beta,m)$ and $C_{E,q}(\mu,\beta,m)$ were dropped off, then the thermodynamic consistency would not be preserved for $\mu > m$.

\subsection{Comparison with previous results in the literature and discussion}
\label{subsec:comparison}

 It is important to notice that the $p$-independent contributions $C_{N,q}$ and $C_{E,q}$, which appear for fermions when $\mu > m$ in $\langle N \rangle$ and $\langle E \rangle$ respectively, have been disregarded in previous works, see e.g.~\cite{CW12,Plastino,Teweldeberhan:2005wq}. These terms are consequence of the discontinuity in the integrand of Eq.~(\ref{partitionfunc}), which appears because of the definition of the q-exponential and q-logarithm functions we are using in Eqs.~(\ref{eq:eq}) and (\ref{eq:logq}). Note that we are considering the same definitions for these functions as the ones used in many previous references, in particular~\cite{Teweldeberhan:2005wq,Plastino}.  The existence of this discontinuity was already pointed out and discussed in~\cite{Teweldeberhan:2005wq} (see Fig.~2 and Eqs.~(25) and (26) in that reference). However, in this and in other works it has not been taken into account the effect coming from the fact that the location of the discontinuity is a function of the 
chemical potential $\mu$, as in the momentum variable the discontinuity appears at  $p = \sqrt{\mu^2 - m^2}$. This means that the discontinuity will lead to some contribution proportional to it when deriving with respect to the chemical potential, which should be added to the contribution from the first derivative of the integrand, as it is written in Eq.~(\ref{eq:Ninte}). Using these arguments, we have corrected some inconsistencies of previous references, and have obtained fully thermodynamical consistent expressions for fermions when $\mu > m$.

It would be possible to get a continuous integrand in Eq.~(\ref{partitionfunc}) by defining the q-log function as $\log_q^{(-)}(x)$ for all values of its argument. However, the motivation to consider different expressions for $\log_q^{(+)}$ and $\log_q^{(-)}$, as well as for $e_q^{(+)}$ and $e_q^{(-)}$, was already stressed in the literature, and it is based on the fact that they fulfill the desirable logarithm-like  $\log_q(x) + \log_q(\frac{1}{x}) = 0$ and exponential-like $e_q(x) e_q(-x) = 1$ relations. See e.g.~\cite{Teweldeberhan:2005wq} for a discussion.

One way to avoid the explicit appearance of the p-independent terms in Eqs.~(\ref{occnumb}) and~(\ref{occE}) is by modifying the definition of the occupation number for $x < 0$, in the form
\begin{equation} 
n_q^{(-)}(p) \to \tilde{n}_q^{(-)}(p) = n_q^{(-)}(p) + f(p,\mu,\beta,m) \,.
\end{equation}
This occupation number contributes only in the interval $\int_0^{\sqrt{\mu^2 - m^2}} dp$ in the integration in momentum, see Eq.~(\ref{eq:Ninte}), and $f(p,\mu,\beta,m)$ is any function which fulfills
\begin{eqnarray}
&&\frac{1}{2\pi^2}\int_0^{\sqrt{\mu^2-m^2}} dp \, p^2 f(p,\mu,\beta,m) = C_{N,q}(\mu,\beta,m) \,, \\
&&\frac{1}{2\pi^2}\int_0^{\sqrt{\mu^2-m^2}} dp \, p^2 E_p f(p,\mu,\beta,m) = C_{E,q}(\mu,\beta,m) \,.
\end{eqnarray}
Then the results of Eqs.~(\ref{occnumb}) and~(\ref{occE}) are
reproduced as it can be easily checked. The point is that
$f(p,\mu,\beta,m)$ is not uniquely determined. There is not a good
criterium to choose one particular expression for $f(p,\mu,\beta,m)$,
so we prefer to leave the expressions for $\langle N\rangle$ and
$\langle E\rangle$ as defined in Eqs.~(\ref{occnumb})
and~(\ref{occE}).

\begin{figure}[tbp]
\begin{center}
\epsfig{figure=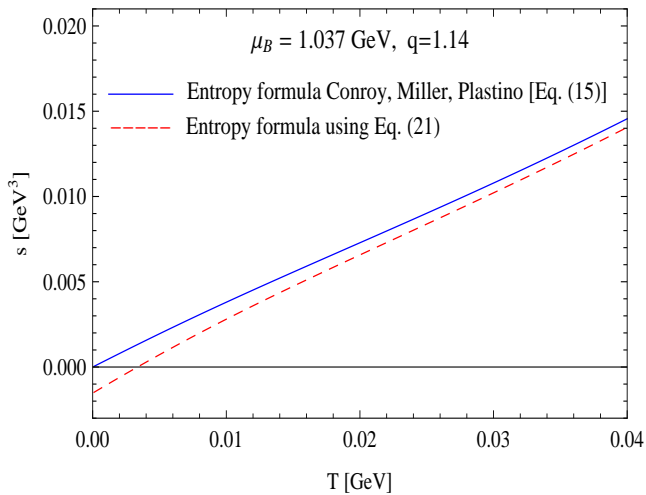,height=6.5cm,width=8.5cm}
\end{center}
\caption{Entropy density $s = S/V = (S^{(+)} + S^{(-)})/V$ as a function of temperature by using the different versions of the entropy formula $S^{(-)}$ in $x < 0$: we plot as continuous (blue) line Eq.~(\ref{eq:S}) which is the same as the Conroy, Miller and Plastino formula~\cite{Plastino}, and as dashed (red) line the result using Eq.~(\ref{eq:Sm}). We include in the computation (anti)protons + (anti)neutrons with baryonic chemical potential $\mu_B = 1.037\,\GeV$ and $q=1.14$ (see section~\ref{sec:Thermo_had_sys} for details).}
\label{fig:entropy}
\end{figure}

The partition function can be obtained also from the total number of particles by integrating in $\mu$, as follows from Eq.~(\ref{eq:avN}). The entropy is then obtained by deriving the partition function with respect to temperature, as in Eq.~(\ref{eq:Sthermo}). When making the computation one can realize that if one considered $q$ instead of $\tilde q$ in the occupation number for $x < 0$, see Eq.~(\ref{eq:nq}), as it is explicitly written in~\cite{Plastino}, the expression of the entropy which is obtained is not the one postulated by these authors, but a much more complicated expression which writes
\begin{equation}
S^{(-)} = V \int \frac{d^3p}{(2\pi)^3} \Theta(-x) \left[ x [\bar{n}_q^{(-)}(x)]^q + \frac{[e_q^{(-)}(x)]^{(1-2q)}}{2q-1} {}_{2}F_{1}[q,2q-1,2q, \xi \cdot [e_q^{(-)}(x)]^{-1}]  \right] \,, \label{eq:Sm}
\end{equation}
where ${}_{2}F_{1}$ is the gauss hypergeometric function.  So, we assume that the presence of $q$ instead of $\tilde q$ in the expression for $\langle N \rangle$ in \cite{Plastino} was a typo, which was wrongly assumed also by~\cite{CW12}.  Only when using $\tilde q$ in the occupation number, Eq.~(\ref{eq:nq}), one gets exactly the Tsallis expression of the entropy written in~\cite{Plastino} (apart from the term $C_{N,q}$ which appears for the reasons discussed above). The entropy of Eq.~(\ref{eq:Sm}) has also the peculiar issue that it leads to negative values in the zero temperature limit, while the entropy of~\cite{Plastino} does not, see Fig.~\ref{fig:entropy}.

On the other hand, it is worth mentioning that the entropy of \cite{CW12} (Eqs.~(21) and (24) in that reference) is not the same as the one in \cite{Plastino} (Eq.~(4) in that reference) for $x < 0$, as it can be easily seen by considering that in the former case they define the q-log function the same for all values of its argument. This means that these authors are not consistent with each other. A computation of the occupation number derived from the entropy of~\cite{CW12} for $x < 0$ leads to the result
\begin{equation}
n_q^{(-)}(x) = -\Bigg(\frac{1}{[e_q^{(-)}(x)]^{-1} - \xi}\Bigg)^{q}\,,
\end{equation}
which is different from the expression of \cite{Plastino} and from the one the authors of \cite{CW12} claim.
 
With these results we conclude that the partition function in
Eq.~(\ref{partitionfunc}) does represent the relevant function for an
ideal quantum gas in Tsallis statistics. Observe that for $\mu=0$ and
$p$ sufficiently high the partition function defined here is similar
to those used in Refs.~\cite{Deppman12,Deppman:2012qt} to extend
Hagedorn's theory to non extensive statistics. Therefore definition
given here is in accordance with the non extensive self-consistent
thermodynamics. In the following we explore some of the features of
the thermodynamical systems described by the partition function
written in Eq.~(\ref{partitionfunc}).
There is an intrinsic gain in the knowledge of the non extensive partition function, since it is closer to the methods of Statistical Mechanics. However, in order to go deeper into more fundamental aspects one needs knowledge about nonperturbative QCD that are not available at present.

\section{Thermodynamical properties of hadronic systems}
\label{sec:Thermo_had_sys}

Before studying the properties of a thermodynamically equilibrated
hadronic system in the non extensive thermodynamics we have to find
the region where this system can exist. In fact, due to the transition
from confined to deconfined regimes, the hadronic matter can be found
only below the phase transition line. There are different proposals for the conditions determining the transition line. Cleymans and Redlich
~\cite{CR_1998,CR} pointed out that the transition line can be determined by the condition that $\langle E \rangle/\langle N \rangle=1 \, \GeV$. This result was obtained through a systematic analysis of particle yields from HEP experiments. In \cite{BMS} the transition line was obtained in terms of the
  total baryon density with the help of the hadron
  resonance gas model. In \cite{MS}, the freeze-out condition was
  determined from an interpolation between a resonance gas used at low
  densities and repulsive nucleonic matter at low temperatures.
In \cite{tawfik_2006}, the author proposed that the chemical
freeze-out of hadrons in heavy-ion collisions is characterized by the
entropy density and its value was  taken from Lattice QCD (LQCD)
calculations at zero chemical potential. In an interesting and more recent analysis \cite{tawfik_2014}, higher order multiplicity
moments obtained with the hadron resonance gas were used in the
calculation of the standard deviation, the variance (or
susceptibility), the skewness and the kurtosis,
quantities which are related to the cumulants \cite{nuxu}, that can be
experimentally determined and are also of interest in LQCD
calculations. The chemical freeze-out curve was then described in
terms of the susceptibility of the system.

In the present work we use two different methods to find the transition line: the condition that $\langle E \rangle/\langle N \rangle=1 \, \GeV$, as in Refs.~\cite{CR_1998,CR}, and the entropy density condition proposed in Ref.~\cite{tawfik_2006}.
 These conditions were compared with a systematic analysis of particle yields from HEP experiments, where the Boltzmann's statistics was used, but since yields are calculated by the integration over all energy states, and considering that the differences between the distributions obtained through Boltzmann statistics and Tsallis statistics are relevant only in the high energy tail, we can assume with some confidence that the same relation holds 
in the non extensive case. Of course this assumption must be checked by a similar analysis of experimental data, now using the non extensive formulas derived here, a task that is  beyond the scope of the present work. However we present below some 
evidences that this hypothesis is correct.

In the following we assume that the entropic index, $q_o$, is a fixed property of the hadronic matter with its value determined in the analysis of $p_T$-distributions and in the study of the hadronic mass spectrum in Ref.~\cite{Lucas}, so we set $q_o=1.14$, although in some cases we analyze the behavior of some quantities for different values of $q$. In the following we refer to Boltzmann-Gibbs statistics also by $q=1$.

\begin{figure}[htb]
\begin{tabular}{cc}
\includegraphics[width=8.cm,angle=0]{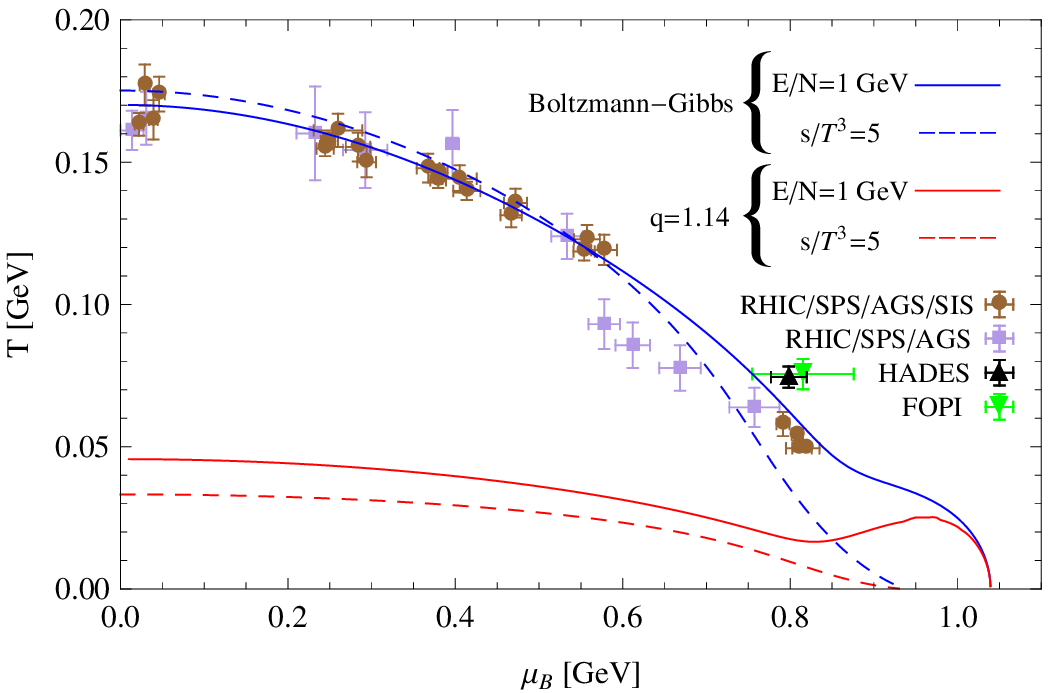} & 
\hspace{0.5cm}\includegraphics[width=8.cm,angle=0]{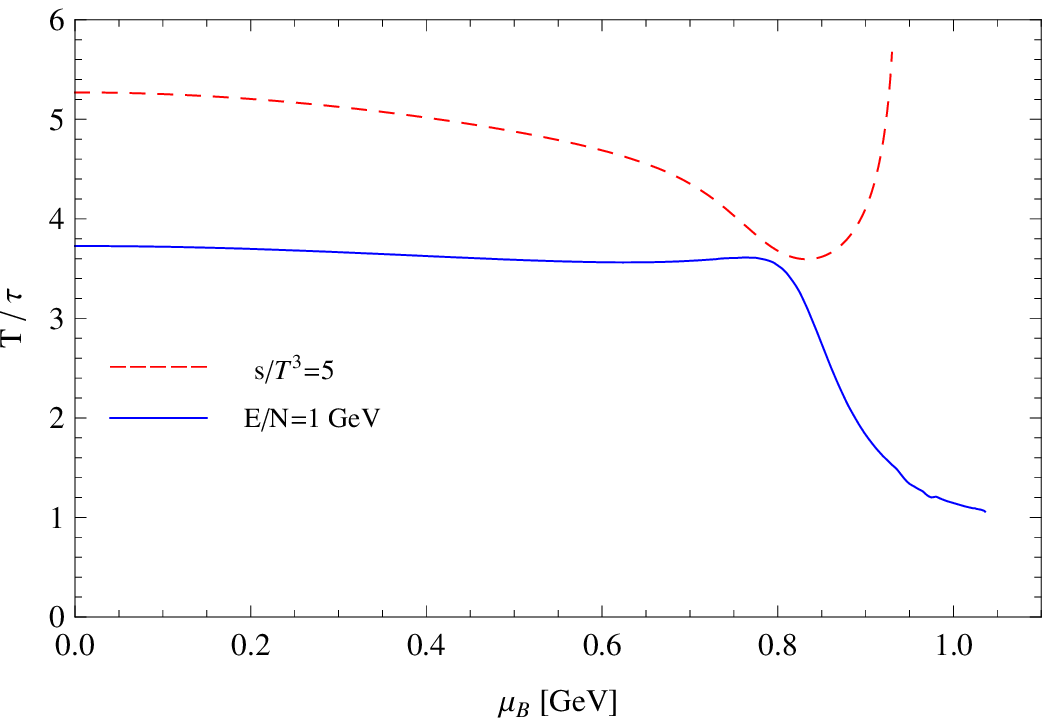}
\end{tabular}
\caption{Left: Chemical freeze-out line $T = T(\mu_B)$ obtained by assuming $\langle E\rangle / \langle N \rangle = 1\, \GeV$ (continuous lines) and $s/T^3 = 5$ (dashed lines). We plot the result by using Boltzmann-Gibbs statistics, and Tsallis statistics with $q=1.14$. Experimental data are taken from: RHIC/SPS/AGS/SIS~\cite{Cleymans:2005xv}, RHIC/SPS/AGS~\cite{Andronic:2005yp}, HADES~\cite{Agakishiev:2010rs}, and FOPI~\cite{Lopez:2007zz}. Right: Quotient between physical temperature $T$ obtained with BG statistics, and the effective temperature $\tau$ obtained with Tsallis statistics and $q=1.14$ according to two different hypotheses: the energy per particle and the entropy density conditions; as a function of the baryonic chemical potential. We consider $\mu = 0$ for mesons in both figures.}
\label{Txmu}
\end{figure}

The system of interest here is a gas composed by different hadronic species in thermodynamical and chemical equilibrium. The partition function is then given by
\begin{equation}
\log \Xi_q(V,T,\{\mu\})=\sum_i \log \Xi_q(V,T,\mu_i)\,, \label{eq:logZ}
\end{equation}
where $\mu_i$ refers to the chemical potential for the {\it i-th}
hadron. The lowest-lying hadrons considered in our calculations are
taken from the Particle Data Group~\cite{Beringer:1900zz}, and some of
them are presented in Tables~\ref{TableMesons} and \ref{TableBaryons}.
The total numbers of hadronic states considered are 808 for mesons and
1168 for baryons ($+$ anti-baryons), including degeneracies, which
correspond to a maximum value of the mass about $11\,\GeV$ and
$5.8\,\GeV$ respectively. The computation will be performed by
restricting the ensemble summation in Eq.~(\ref{eq:logZ}) to zero
strangeness (see~\cite{CR} for details).  We will focus first on a
study of the phase transition by assuming $\mu=0$ for all mesons and
considering that all baryons have the same chemical potential value
$\mu_B \neq 0$. After that, the effect of a nonzero value of the
chemical potential for pions will be explored as well.  We have not
compared our results with LQCD data because these data are available
only up to baryon chemical potentials much lower than the ones we need
to cover the whole $T-\mu_B$ plane \cite{endrodi}.  In \cite{ejiri},
for instance $\mu/T$ reaches 2, what means that $\mu$ is of the order
of $400-500$ MeV corresponding to $\mu_B$ of the order of $133 - 166$
MeV. In addition, one can observe in Fig.~\ref{Txmu} that for small
chemical potential the effective temperature remains practically
unchanged, so the theoretical results would not differ sensitively
from those obtained with the usual Boltzmann-Gibbs statistics after
the temperature scalability observed in this work is taken into
account. For the case of null chemical potential a comparison with
LQCD was already done in Ref.~\cite{Deppman:2012qt}.

The phase transition line obtained  with the energy per particle condition, as described above, is reported in Fig.~\ref{Txmu} (left), and it is compared to the Boltzmann's result, corresponding to $q\rightarrow 1$. For the sake of clarity we refer to the temperature obtained with BG statistics as $T$ and to the temperature obtained with Tsallis statistics as $\tau$. The relation between $\tau$ and $T$ was already investigated in Ref.~\cite{Deppman:2012qt}, and it was found that for a fixed chemical potential there is a linear correspondence between both quantities. The relation between both quantities when varying $\mu_B$ is shown in Fig.~\ref{Txmu} (right). For the sake of comparison the transition line obtained through the entropy condition was calculated, and its results are shown in Fig.~\ref{Txmu}. We observe that both conditions lead to transition lines that are in agreement with the available experimental data. The main differences are in the high chemical potential region, where the entropy 
condition leads to lower temperature, approaching zero around $\mu_B=0.9$, while the energy per particle condition gives higher temperatures in the same region.

\begin{table}
\centering
\renewcommand{\arraystretch}{1.5}
{
\begin{tabular}{|c|c|c|c||c|c|c|c|c|}
\hline
Mesons & Mass & $S$  & $\;\; g \;\;$   &  Mesons      & Mass &$S$ & $\;\; g \;\;$ \\
\hline
$\pi^0$          & 134.98    & $0$  & $1$  & $K^+$ & 493.68    &  $1$  & $1$      \\
$\pi^{+},\pi^{-}$ & 139.57    & $0$  & $2$  & $K^0$     &  497.67   &  1    & 1     \\
$\eta$          & 547.3      & $0$  & $1$  & $K^*(892)^+$ &  891.66   &  1    & 3     \\ 
$\rho(770)$      & 771.1     & $0$ & $9$   & $K^*(892)^0$ &  896.1     &  1    &  3      \\
$\omega(782)$   & 782.57    & $0$  & $3$   & $K_1(1270)$ & 1273      &  1    &  6      \\
$\eta^\prime(958)$ & 957.78  &  $0$ & $1$  & $K_1(1400)$  & 1402       &  1    & 6    \\
$f_0(980)$      & 980        &  $0$ & $1$  & $K_0^*(1430)$ & 1412      &  1    & 2       \\
$a_0(980)$      & 984.7     &  $0$ & $3$  & $K^*(1410)$   & 1414      &  1    & 6     \\
$\phi(1020)$    &  1019.46 &  $0$ & $3$  & $K^*_2(1430)^+$  & 1425.6    &  1    & 5   \\
$h_1(1170)$      & 1170     &  $0$ & $3$  & $K^*_2(1430)^0$  & 1432.4    &  1    & 5  \\
$b_1(1235)$      & 1229.5   &  $0$ & $9$  & $K(1460)$    & 1460      &  1  & 10     \\  
$\cdots$         & $\cdots$ &  $\cdots$ & $\cdots$  & $\cdots$   & $\cdots$  &  $\cdots$  & $\cdots$     \\            
\hline 
\end{tabular}
}
\caption{List of the lowest-lying mesons used in Eq.~(\ref{eq:logZ}). We include some of their properties: mass, strangeness ($S$) and degeneracy ($g$).  The corresponding anti-mesons with $S=-1$ are not shown in the table but they are considered as well in the computation. The dots indicate that heavier mesons are included in the computation, although they are not explicitly shown in this table due to lack of space.}\label{TableMesons}
\end{table}

\begin{table}
\centering
\renewcommand{\arraystretch}{1.5}
{
\begin{tabular}{|c|c|c|c||c|c|c|c|}
\hline
Baryons      & Mass      & $S$ &  $\;\;g\;\;$  & Baryons   & Mass  & $S$     &  $\;\;g\;\;$ \\
\hline
$p$   &  938.27  &  $0$ &  2        & $\Lambda^0$   &  1115.68  &  $-1$     &  2    \\
$n$ &  939.56  &  $0$ &  2          & $\Sigma^+$    &  1189.37  &  $-1$     &  2    \\
$\Delta(1232)$   &  1232    &  $0$ &  16        & $\Sigma^0$    &  1192.64  &  $-1$     &  2    \\
$N(1440)$    &  1440    &  $0$ &  4 & $\Sigma^-$    &  1197.45  &  $-1$     &  2    \\
$N(1520)$    &  1520    &  $0$ &  8 & $\Sigma^{*+}$ &  1382.8   &  $-1$     &  4    \\
$N(1535)$    &  1535    &  $0$ &  4 & $\Sigma^{*0}$ &  1383.7   &  $-1$     &  4    \\
$\Delta(1600)$&  1600    &  $0$ &  16& $\Sigma^{*-}$ &  1387.2   &  $-1$     &  4    \\
$\Delta(1620)$&  1620    &  $0$ &  8 & $\Lambda(1405)$            &  1406     &  $-1$     &  2    \\
$\cdots$     & $\cdots$ &  $\cdots$ & $\cdots$  & $\cdots$   & $\cdots$  &  $\cdots$  & $\cdots$     \\            
\hline
$\Xi^0$      &  1314.83 &  $-2$ &   2  & $\Omega^-$    &  1672.45  &  $-3$     &  4    \\
$\Xi^-$      &  1321.31 &  $-2$ &   2     & & & &       \\
$\Xi^{*0}$   &  1531.8  &  $-2$ &   4     &&&&      \\
$\Xi^{*-}$   &  1535    &  $-2$ &   4     &&&&        \\
$\cdots$     & $\cdots$ &  $\cdots$ & $\cdots$  &   &&& \\
\hline
\end{tabular}
}
\caption{List of the lowest-lying baryons used in Eq.~(\ref{eq:logZ}).
We show only the baryons with baryonic number $B=1$. The corresponding anti-baryons $B=-1$ and $S=0, 1, 2$ and $3$ are considered as well in the computation. 
}\label{TableBaryons}
\end{table}

\begin{figure}[htb]
\includegraphics[width=8.cm,angle=0]{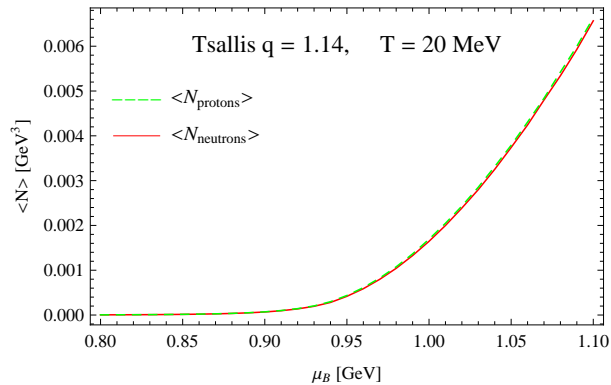}
\caption{Density of protons and neutrons as a function of the baryonic chemical potential, in Tsallis statistics with $q=1.14$. 
We consider $T = 20\, \MeV$. Dashed green line corresponds to the density of protons, while continuous red line is the density of neutrons.}
\label{fig:Npn}
\end{figure}

We observe in Fig.~\ref{Txmu} (left) that for both $q=1$ and $q=1.14$
we obtain lines with similar shapes, but $\tau$ is always lower than
$T$. The ratio between both temperatures is practically constant up to
$\mu_B \simeq 0.8\, \GeV$, as can be seen in Fig.~\ref{Txmu} (right),
and a consequence is that the curve which results from the
multiplication of the results with $q=1.14$ by a constant factor equal
to 3.65 (in the energy per particle case) almost reproduces the result
with BG statistics in Fig.~\ref{Txmu} (left). This constant factor is
of the order of 5 when using the entropy criterium.  In the following
we will restrict our discussion to the results obtained with the
energy per particle hypothesis. In this case the chemical freeze-out
lines spam over the region of $0<\mu_B<1039.2\,\MeV$ with the same
maximum value for $\mu_B$, corresponding to a null critical
temperature. Near this maximum value the relation $T/\tau$ tends to be
close to $1$. The curves in Fig.~\ref{Txmu} (left) show an inflection
for $\mu_B \sim 0.9 \, \GeV$ which is related to the sharp increase in
the baryon density as the baryonic chemical potential approaches the
proton/neutron mass $m_{p,n} \simeq 0.94 \, \GeV$. This increase is
displayed in Fig.~\ref{fig:Npn}, where we plot the proton and neutron
densities as a function of $\mu_B$ in Tsallis statistics. A similar
behavior is observed in Boltzmann-Gibbs statistics.

For $\mu_B=0$ the effective temperature is $T_o=45.6\,\MeV$ for
$q=1.14$, which is not in agreement with the value $T_o = (60.7 \pm
0.5) \, \MeV$ found in the analysis of the
$p_T$-distributions~\cite{Lucas}. This disagreement can be related to
the value adopted for $\langle E \rangle/\langle N \rangle$, which
still must be checked by analysis of experimental data with the non
extensive statistic.  In order to provide an estimate of the sensitivity of the value of $T_o$ when changing $q$ it is worth mentioning that the effective temperature for a slightly smaller value of the entropic index, $q=1.12$, is $T_o=61.0\, \MeV$, which is in agreement with that reference.  

The transition line determines the region where
the confined states exist (below the line), and the region where one
expects to find the quark-gluon plasma (above the line).

\begin{figure}[htb]
\begin{tabular}{cc}
\includegraphics[width=8.cm,angle=0]{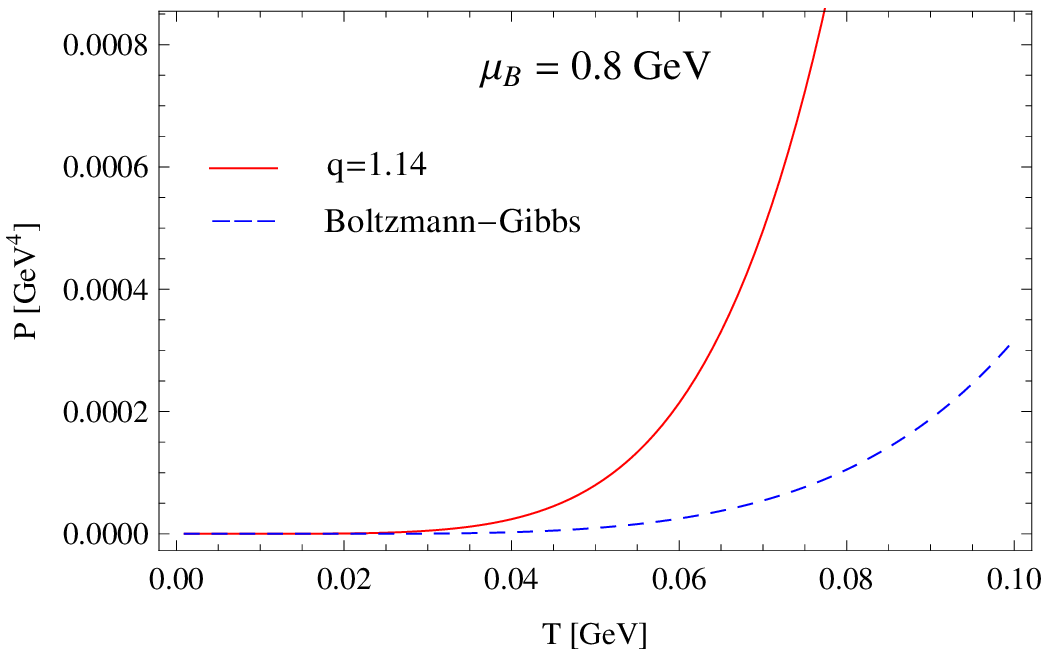} & 
\includegraphics[width=8.cm,angle=0]{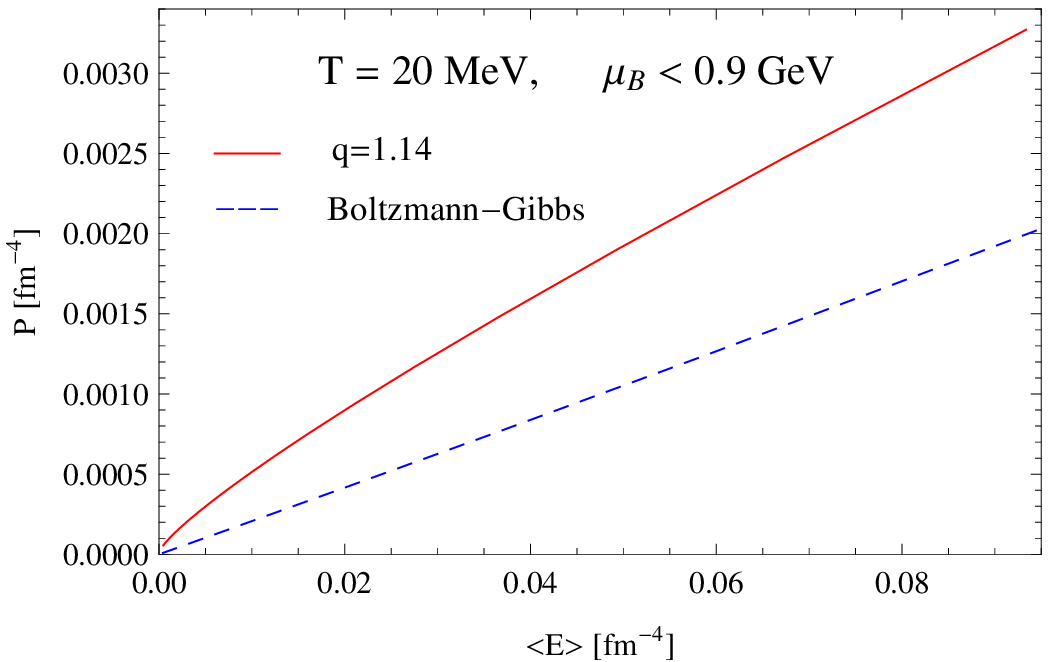}
\end{tabular}
\caption{Left: Pressure as a function of temperature. The baryonic chemical potential is kept fixed to the value $\mu_B = 0.8 \, \GeV$. Right: Pressure as a function of energy density when changing the baryonic chemical potential in the range $0 \, \GeV < \mu_B < 0.9 \, \GeV $. The temperature is kept fixed to the value $T = 20\, \MeV$. In both figures we plot as a dashed blue line the result in Boltzmann-Gibbs statistics, and as a continuous red line the result in Tsallis statistics with $q=1.14$.}
\label{PxT}
\end{figure}

\begin{figure}[htb]
\begin{tabular}{cc}
\includegraphics[width=8.cm,angle=0]{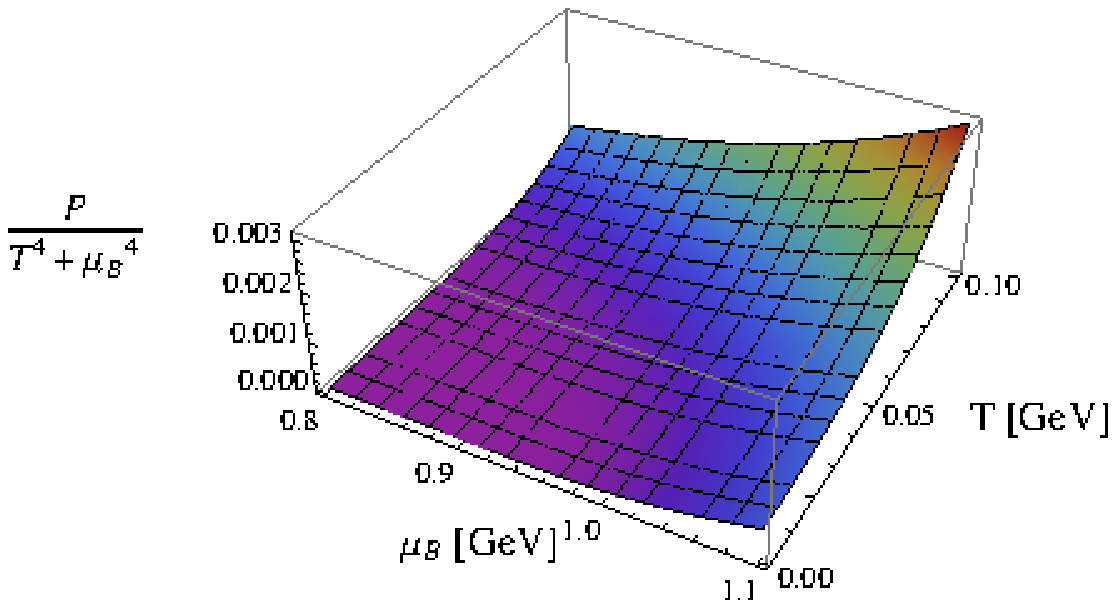} & 
\includegraphics[width=8.cm,angle=0]{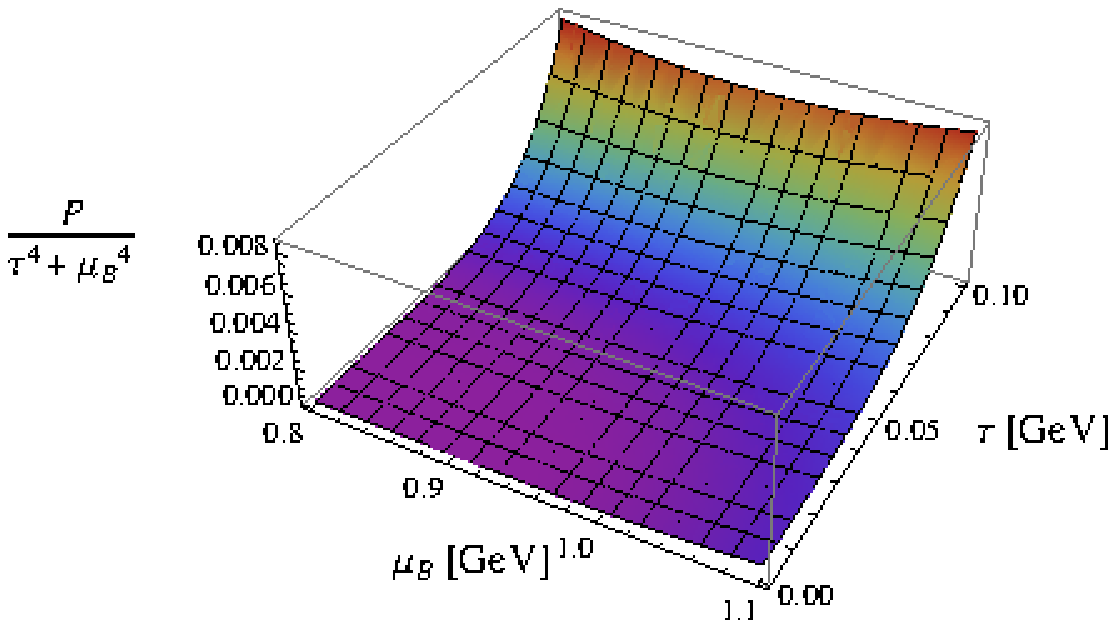}
\end{tabular}
\caption{Pressure (normalized to $T^4 + \mu_B^4$) as a function of temperature and baryonic chemical potential. Left: Result using Boltzmann-Gibbs statistics. Right: Result using Tsallis statistics with $q=1.14$.}
\label{P3D}
\end{figure}

\begin{figure}[htb]
\begin{tabular}{cc}
\includegraphics[width=8.cm,angle=0]{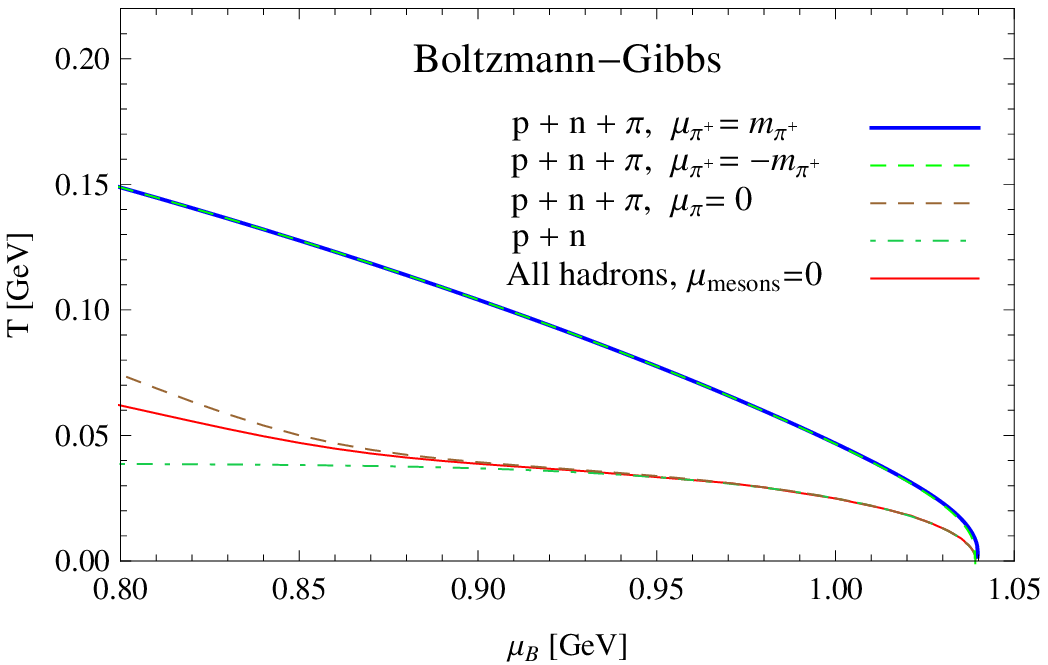} &
\includegraphics[width=8.cm,angle=0]{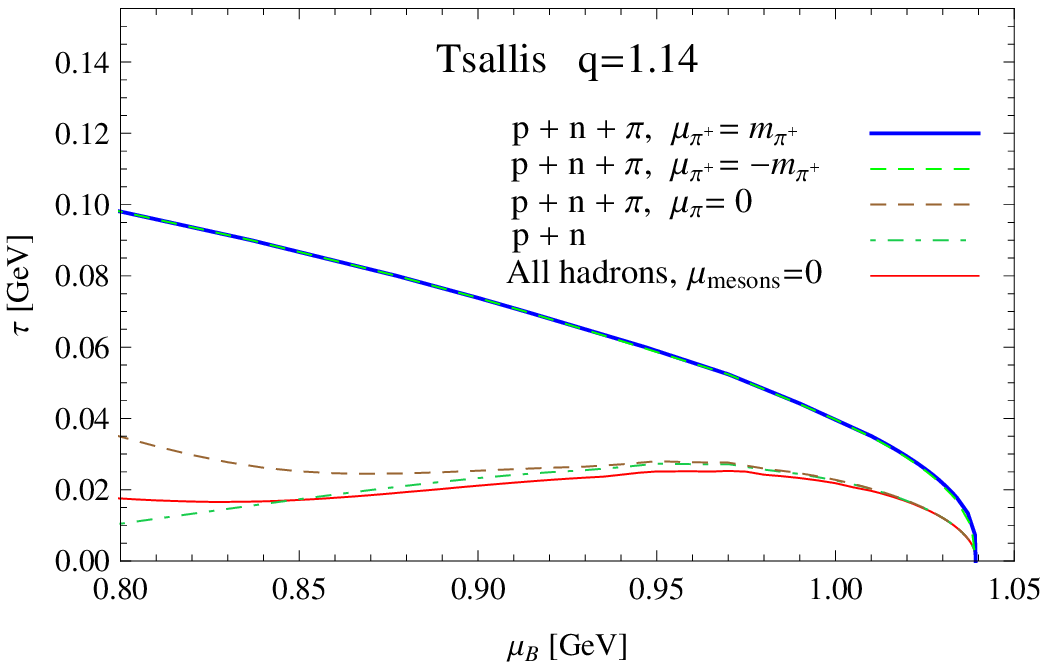}
\end{tabular}
\caption{Chemical freeze-out line $T = T(\mu_B)$ obtained by assuming $\langle E\rangle / \langle N \rangle = 1\, \GeV$. Left: Boltzmann-Gibbs statistics. Right: Tsallis statistics with $q=1.14$. We show the result in different cases, including: i) ``(anti)protons + (anti)neutrons'', ii) ``(anti)protons + (anti)neutrons + pions'' with $\mu_\pi = 0$, $\mu_{\pi^+} = m_{\pi^+}$ or $\mu_{\pi^+} = - m_{\pi^+}$, iii) all hadrons with $\mu_{\textrm{mesons}}=0$ (as in Fig.~\ref{Txmu}). In ii) we consider $\mu_{n} = \mu_{p} - \mu_{\pi^+}$, and in this case it is represented in the horizontal axis $\mu_{p}$ when $\mu_{\pi^+} = m_{\pi^+}$, and $\mu_{n}$ when $\mu_{\pi^+} = -m_{\pi^+}$. }
\label{Txmumu}
\end{figure}

\begin{figure}[htb]
\begin{tabular}{cc}
\includegraphics[width=8.cm,angle=0]{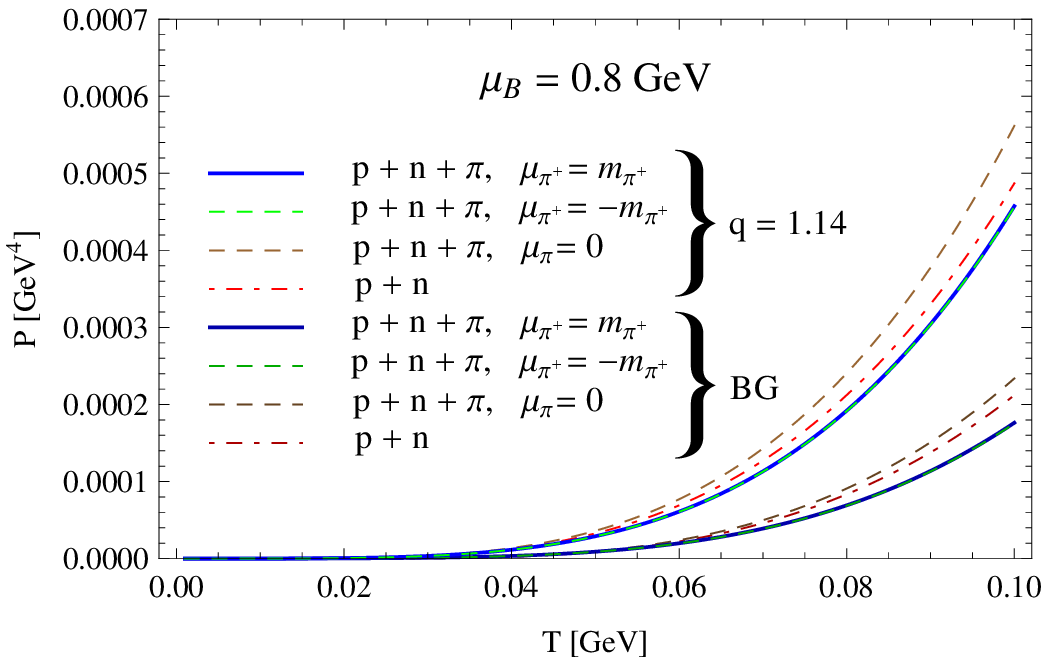} &
\includegraphics[width=8.cm,angle=0]{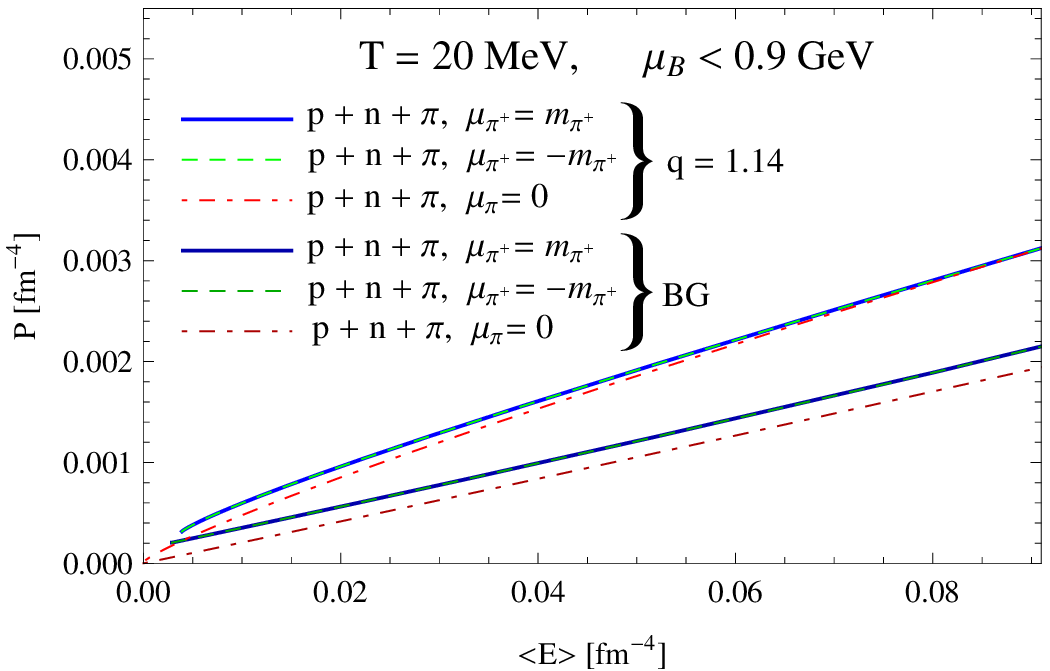}
\end{tabular}
\caption{The same as in Fig.~\ref{PxT}, but including only the lowest states in the hadron spectrum, i.e. ``(anti)protons + (anti)neutrons'' and ``(anti)protons + (anti)neutrons + pions''. In the latter case, we plot curves with either $\mu_\pi = 0$, $\mu_{\pi^+} = m_{\pi^+}$ and $\mu_{\pi^+} = - m_{\pi^+}$. For the clarity of the figure it is not plotted the result for p + n in the right figure, as it is very close to the result with p + n + $\pi$, $\mu_{\pi} = 0$. We display the results using Boltzmann-Gibbs (lower curves) and Tsallis statistics with $q=1.14$ (higher curves). See Figs.~\ref{PxT} and \ref{Txmumu} for other details.}
\label{fig:PmuBmu}
\end{figure}

In Fig.~\ref{PxT} (left) we plot the pressure as a function of the
temperature for the chemical potentials fixed at the value
$\mu_B=0.8\,\GeV$ for the baryonic chemical potential and $\mu=0$ for
all mesons with $q=1$ and $q=1.14$.  We observe that the pressure
increases faster in the non extensive case, $q=1.14$, than in the
extensive one, $q=1$.  In Fig.~\ref{PxT} (right) we show the results for the
pressure as a function of the energy density for $T=20\,\MeV$. Also
here we observe that the pressure increases faster in the case of
$q=1.14$ in comparison with the case $q=1$. We show in Fig.~\ref{P3D} the pressure as a function of the temperature and chemical potential in the regime $0.8 \, \GeV < \mu_B < 1.1 \, \GeV$ and $0 \, \GeV< T < 0.1 \, \GeV$, in Boltzmann-Gibbs (left) and in Tsallis with $q=1.14$ (right). The more rapid growth of the pressure in Tsallis statistics in comparison with Boltzmann-Gibbs is clear for all the values of $\mu_B$ displayed, but this effect becomes much stronger for $\mu_B \sim 0.8 \, \GeV$, see also Fig.~\ref{PxT} (left). From Figs.~\ref{PxT} and \ref{P3D} one can conclude that the EOS for hadronic matter obtained with the Tsallis statistics is harder than the one obtained from the BG case.

Up to now we have studied the equation of state including the spectrum of hadrons in Table~\ref{TableMesons} and \ref{TableBaryons}. It would be interesting to analyze also the case in which only protons, neutrons and possibly pions contribute to the equation of state, all of them with nonzero chemical potential. We have studied the finite pion chemical potential in two different cases: a) $\mu_{\pi^+} = m_{\pi^+} = -\mu_{\pi^-} $, b)  $\mu_{\pi^+} = -m_{\pi^+} = -\mu_{\pi^-} $; and $\mu_{\pi^0} = 0$, $\mu_n = \mu_p - \mu_{\pi^+}$ in both cases. These values are relevant for the study of protoneutron stars~\cite{Menezes:2014wqa}.
The phase transition lines in Boltzmann-Gibbs and in Tsallis statistics in the regime of high baryonic chemical potential are plotted in Fig.~\ref{Txmumu}. Note that the effect of finite pion chemical potential is to increase the transition line to higher values of temperature, and the curves with $\mu_{\pi^+} = m_{\pi^+}$ and $\mu_{\pi^+} = - m_{\pi^+}$ coincide, as we have not introduced electrons in 
the computation. 
We show in Fig.~\ref{fig:PmuBmu} the result for the equation of state considering zero and finite pion chemical potential. The effect of pions is to increase the values for the pressure with respect to the case with only protons and neutrons. When considering a nonzero value for the pion chemical potential this leads to a noticeable effect on the EOS, as it becomes harder either in Boltzmann-Gibbs or in Tsallis statistics, see Fig.~\ref{fig:PmuBmu} (right).

\section{Conclusions}
\label{sec:conclusions}

In this work we developed the non extensive thermodynamics for an
ideal quantum gas for both bosons and fermions from the partition
function defined here for the first time. Then we showed that the
partition function and the thermodynamics derived from it is
equivalent to the thermodynamics derived from the entropy proposed by
Conroy, Miller and Plastino~\cite{Plastino}, and also by Cleymans and
Worku~\cite{CW12} when $\mu \le m$. For $\mu > m$ and for fermions some
inconsistencies of previous references were addressed, and our result
is fully thermodynamically consistent.  In the limit of high energies,
the partition function is in accordance with that proposed in
Ref.~\cite{Deppman12}, and thus the partition function defined here is
self-consistent in the sense proposed by Hagedorn.

Some thermodynamical functions are derived from the partition function
for hadronic systems with different values of chemical
potentials. Particularly we analyze how pressure and energy densities
vary when the entropic index or the chemical potentials vary, and we
obtain the chemical freeze-out line by using two different
hypotheses. A discussion about a discontinuity observed in the first
derivatives of the partition function is done for the first time.

The results presented in this work can be applied to stellar matter,
where high pressures are necessary to compensate for the gravitational
force so that protoneutron star stability and correct macroscopic
properties are attained. This study is performed in other work~\cite{Menezes:2014wqa}, and the results compared with the ones existing in the literature~\cite{lavagno}.

\subsection*{Acknowledgments}
This work has been supported by Plan Nacional de Altas Energ\'{\i}as
(FPA2011-25948), Junta de Andaluc{\'\i}a grant FQM-225, Generalitat de
Catalunya grant 2014-SGR-1450, Spanish MINECO's Consolider-Ingenio
2010 Programme CPAN (CSD2007-00042), Centro de Excelencia Severo Ochoa
Programme grant SEV-2012-0234, by CNPq (Brazil) and FAPESC (Brazil)
under project 2716/2012,TR 2012000344, and FAPESP (Brazil) under grant
2013/24468-1. A.D. acknowledges the support from CNPq under grant
305639/2010-2.  E.M.  would like to thank the Instituto de F\'{\i}sica
of the Universidade de S\~ao Paulo for their hospitality and support
during the completion of parts of this work. The research of E.M. is
supported by the Juan de la Cierva Program of the Spanish MINECO grant JCI-2010-06699.

\bibliographystyle{aipproc}

\end{document}